\documentclass{emulateapj}

\usepackage{natbib}
\usepackage{amsmath}

\newcommand{\bvri}{\protect\hbox{$BV\!RI$} }

\newcommand{\about}{$\sim\!\!$~}

\newcommand{\err}[2]{\ensuremath{^{+#1}_{-#2}}}

\def\lsim{\hbox{\rlap{\raise 0.425ex\hbox{$<$}}\lower 0.65ex\hbox{$\sim$}}}
\def\gsim{\hbox{\rlap{\raise 0.425ex\hbox{$>$}}\lower 0.65ex\hbox{$\sim$}}}

\def\arcsec{\hbox{$^{\prime\prime}$}}
\def\arcdeg{\mbox{$^\circ$}}

\shorttitle{Progenitor System or Stellar Remnant of SN~2008ha}
\shortauthors{Foley et~al.}

\begin{document}

\title{Possible Detection of the Stellar Donor or Remnant for the\\ Type~I\lowercase{ax} Supernova 2008\lowercase{ha}}

\def\illast{1}
\def\illphy{2}
\def\rut{3}
\def\ucsb{4}
\def\kitp{5}
\def\cfa{6}
\def\noao{7}
\def\stsci{8}
\def\aar{9}

\author{
{Ryan~J.~Foley}\altaffilmark{\illast,\illphy},
{Curtis~McCully}\altaffilmark{\rut},
{Saurabh~W.~Jha}\altaffilmark{\rut},
{Lars~Bildsten}\altaffilmark{\kitp,\ucsb},
{Wen-fai~Fong}\altaffilmark{\cfa},
{Gautham~Narayan}\altaffilmark{\noao},
{Armin~Rest}\altaffilmark{\noao}, and
{Maximilian~D.~Stritzinger}\altaffilmark{\aar}
}

\altaffiltext{\illast}{
Astronomy Department,
University of Illinois at Urbana-Champaign,
1002 W.\ Green Street,
Urbana, IL 61801, USA
}
\altaffiltext{\illphy}{
Department of Physics,
University of Illinois Urbana-Champaign,
1110 W.\ Green Street,
Urbana, IL 61801, USA
}
\altaffiltext{\rut}{
Department of Physics and Astronomy,
Rutgers, The State University of New Jersey,
136 Frelinghuysen Road,
Piscataway, NJ 08854, USA
}
\altaffiltext{\kitp}{
Kavli Institute for Theoretical Physics,
Santa Barbara, CA 93106, USA
}
\altaffiltext{\ucsb}{
Department of Physics,
University of California,
Santa Barbara, CA 93106, USA
}
\altaffiltext{\cfa}{
Harvard-Smithsonian Center for Astrophysics,
60 Garden Street, 
Cambridge, MA 02138, USA
}
\altaffiltext{\noao}{
National Optical Astronomy Observatory,
950 North Cherry Avenue,
Tucson, AZ 85719-4933, USA
}
\altaffiltext{\stsci}{
Space Telescope Science Institute,
3700 San Martin Drive,
Baltimore, MD 21218, USA
}
\altaffiltext{\aar}{
Department of Physics and Astronomy,
Aarhus University,
Ny Munkegade,
DK-8000 Aarhus C, Denmark
}

\begin{abstract}
  Type Iax supernovae (SNe~Iax) are thermonuclear explosions that are
  related to SNe~Ia, but are physically distinct.  The most important
  differences are that SNe~Iax have significantly lower luminosity
  (1\%--50\% that of typical SNe~Ia), lower ejecta mass (\about0.1 --
  0.5~M$_{\sun}$), and may leave a bound remnant.  The most extreme
  SN~Iax is SN~2008ha, which peaked at $M_{V} = -14.2$~mag, about
  5~mag below that of typical SNe~Ia.  Here, we present {\it Hubble
    Space Telescope} ({\it HST}) images of UGC~12682, the host galaxy
  of SN~2008ha, taken 4.1~years after the peak brightness of
  SN~2008ha.  In these deep, high-resolution images, we detect a
  source coincident (0.86 {\it HST} pixels; 0.043\arcsec;
  1.1~$\sigma$) with the position of SN~2008ha with $M_{\rm F814W} =
  -5.4$~mag.  We determine that this source is unlikely to be a chance
  coincidence, but that scenario cannot be completely ruled out.  If
  this source is directly related to SN~2008ha, it is either the
  luminous bound remnant of the progenitor white dwarf or its
  companion star.  The source is consistent with being an evolved
  $>$3~M$_{\sun}$ initial mass star, and is significantly redder than
  the SN~Iax~2012Z progenitor system, the first detected progenitor
  system for a thermonuclear SN.  If this source is the companion star
  for SN~2008ha, there is a diversity in SN~Iax progenitor systems,
  perhaps related to the diversity in SN~Iax explosions.  If the
  source is the bound remnant of the white dwarf, it must have
  expanded significantly.  Regardless of the nature of this source, we
  constrain the progenitor system of SN~2008ha to have an age of
  $<$80~Myr.
\end{abstract}

\keywords{galaxies---individual(UGC~12682), supernovae---general,
  supernovae---individual (SN~2008ha)}


\defcitealias{McCully14:12z}{M14}

\section{Introduction}\label{s:intro}

Type Iax supernovae (SNe~Iax) are a class of thermonuclear stellar
explosions physically distinct from SNe~Ia, but sharing several
characteristics (see \citealt{Jha06:02cx} and \citealt{Foley13:iax}
for reviews).  This class is the most common of all types of peculiar
SNe by both number and rate, occurring at roughly 30\% the rate of
SNe~Ia \citep{Foley13:iax}.  All observational properties indicate
that these explosions are less energetic than SNe~Ia
\citep[e.g.,][]{Li03:02cx, Phillips07, Foley13:iax}.

While the progenitor systems of SNe~Ia are still uncertain (see
\citealt{Maoz13} for a review), a reasonable progenitor model and
explosion mechanism that explains most SN~Iax observations has
emerged.  As the number of observations has increased
\citep{Foley09:08ha, Foley10:08ha, Foley10:08ge, Foley13:iax,
  Valenti09, Maund10:05hk, McClelland10, Milne10, Narayan11, Kromer13,
  McCully14:iax, Stritzinger14}, the best model for these SNe continues to
be a deflagration of a C/O white dwarf (WD) that does not burn through
the entire WD as first proposed by \citet{Foley09:08ha}.  Because of
He seen in the spectra of two SNe~Iax and their implied delay time of
$\lesssim$500~Myr, \citet{Foley13:iax} expanded on this initial model
by employing a He-star companion, similar to a previously suggested
progenitor system for SNe~Ia \citep{Iben91, Liu10}.  Observations
suggest, and this model implies, that at least some SNe~Iax do not
unbind their progenitor star, leaving behind a remnant
\citep{Foley09:08ha, Foley10:08ha, Foley13:iax, Jordan12, Kromer13,
  McCully14:iax}.  This model can explain all observational properties of
SNe~Iax except perhaps the low polarization seen in a single SN~Iax
\citep{Chornock06, Maund10:05hk}.

\begin{figure*}
\begin{center}
\epsscale{1.15}
\rotatebox{0}{
\plotone{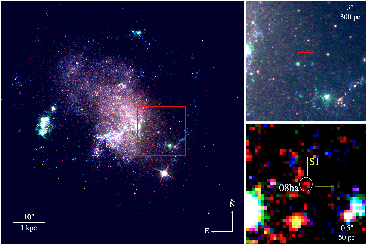}}
\caption{False-color {\it HST} image of UGC~12682, the host galaxy of
  SN~2008ha.  The red, green, and blue channels correspond to F814W,
  F625W, and F435W, respectively.  {\it Left}:
  75\arcsec$\times$75\arcsec\ image.  The red box is
  15\arcsec$\times$15\arcsec\ and centered on the position of
  SN~2008ha.  {\it Top-Right}: 15\arcsec$\times$15\arcsec\ image
  corresponding to the red box in the left panel.  The red box is
  2\arcsec$\times$2\arcsec.  {\it Bottom-Right}:
  2\arcsec$\times$2\arcsec\ image corresponding to the red box in the
  top-right panel.  A 3~$\sigma$ ellipse is centered on the position
  of SN~2008ha.  There is a single source near the center of the
  ellipse, which we label ``S1.''  The bright source to the south-east
  of S1 is an \ion{H}{2} region with a measured metallicity of $12 +
  \log [{\rm O/H}] = 8.16$ \citep{Foley09:08ha}.}\label{f:finder}
\end{center}
\end{figure*}

This model has been directly tested by pre-explosion {\it Hubble Space
  Telescope} ({\it HST}) images of SN~2008ge \citep{Foley10:08ge} and
SN~2012Z (McCully et~al., in press; hereafter
\citetalias{McCully14:12z}).  No progenitor was detected in the
relatively shallow images for SN~2008ge, but the data excluded most
massive stars as the progenitor (as suggested by \citealt{Valenti09}
and \citealt{Moriya10}).  Deep limits on the star formation for the
host galaxy of SN~2008ge also indicated that SN~2008ge did not have a
massive progenitor star \citep{Foley10:08ge}.

\citetalias{McCully14:12z} detected a star in pre-explosion {\it HST}
images at the position of SN~2012Z.  This was the first direct
detection of a thermonuclear SN progenitor system.  This relatively
luminous star is most consistent with a massive ($M \approx
2$~M$_{\sun}$ at the time of explosion) He star, as predicted by
\citet{Foley13:iax}.

In this manuscript, we present post-explosion {\it HST} images of the
host galaxy of SN~2008ha, the least luminous SN~Iax
\citep{Foley09:08ha, Foley10:08ha, Valenti09, Stritzinger14}.
SN~2008ha was discovered on 2008 November 7.17 (UT dates are used
throughout this paper; \citealt{Puckett08}) in UGC~12682, an irregular
galaxy with a recession velocity of 1393~km~s$^{-1}$, corresponding to
a Virgo-infall corrected distance modulus of $\mu = 31.64 \pm
0.15$~mag and $D = 21.3 \pm 1.5$~Mpc, assuming $H_{0} = 73 \pm
5$~km~s$^{-1}$~Mpc$^{-1}$.  \citet{Foley08:08ha} first classified
SN~2008ha as a SN~Iax, but also noted its extreme characteristics and
the possibility that SN~2008ha may not have unbound its progenitor
star.

In the post-explosion images, we detect a point source near the
position of SN~2008ha.  We outline the data acquisition and reduction
in Section~\ref{s:obs}.  In Section~\ref{s:anal}, we detail the
properties of this source, and examine several possibilities for its
nature including being a chance coincidence, SN emission, the stellar
remnant, and the companion star from the progenitor system.  We also
constrain the properties of the progenitor system and remnant if the
source is unrelated to SN~2008ha.  We discuss the implications of our
findings and summarize our conclusions in Section~\ref{s:disc}.


\section{Observations and Data Reduction}\label{s:obs}

UGC~12682 was observed with {\it HST}/ACS on 2012 January 2.2 (Program
GO-12999; PI Foley).  The full observing sequence obtained 1164, 764,
840, and 1240 seconds in the F435W, F555W, F625W, and F814W filters
(corresponding roughly to \bvri\!\!).

We combined exposures (including cosmic ray rejection) using
AstroDrizzle.  We register the individual flatfielded (flt) frames
using TweakReg from the DrizzlePac package.  We registered the
absolute astrometry to ground-based images of the same field using the
F555W exposures.  We then tied the relative astrometry of the other
filters to this image.  The typical rms of the relative astrometry
solution was $0.005\arcsec = 0.1$ ACS pixels.  We drizzle the images
to the native scale of ACS, 0.05\arcsec\ per pixel.  A false-color
image of UGC~12682 and zoomed panels of the region near the position
of SN~2008ha are presented in Figure~\ref{f:finder}.

We photometered the {\it HST} images using the PSF photometry software
DolPhot, an extension of HSTPhot \citep{Dolphin00}.  DolPhot
photometers individual flt frames and then combines the photometry.
We use the suggested parameters for ACS from the Dolphot manual.  We
only consider photometry with a ``flag'' of 0 and an ``object type''
of 1 (point sources) in our results.

Using an $r$-band image taken with the 1.0-m Swope telescope, we were
able to determine the astrometric position of SN~2008ha in the {\it
  HST} images.  The Swope image, from which the \citet{Stritzinger14}
SN~2008ha photometry was derived, had a template image subtracted for
the region around SN~2008ha.  This provides a precise measurement of
the position of SN~2008ha relative to field stars without
contamination from underlying emission (particularly from a nearby
\ion{H}{2} region).

The Swope and {\it HST} images had a total of 14 stars in common with
well-measured positions.  Following the procedures of \citet[and
references therein]{Foley10:08ge}, we determined a geometric
transformation between the two images.  However, tests showed that
inclusion of stars far from SN~2008ha adversely affected the accuracy
of the transformation as determined by looking at positions of stars
not included in the fitting.  The poor accuracy is likely because of
higher-order distortions on large spatial scales that could not be
adequately removed with the limited number of stars.  However, we
found that using only the 8 stars closest to SN~2008ha removed all
obvious bias near the location of SN~2008ha.

Using these 8 stars, the accuracy of the geometric transformation is
robust near the SN position.  However, the nominal uncertainties in
the SN position (\about 0.4 ACS pixels; \about 0.02\arcsec) may be
underestimated given the previously determined errors.  To provide a
better estimate of the positional uncertainty, we re-determined the
position of SN~2008ha 100 times by selecting bootstrapped samples
(with replacement) from the initial sample of 8 stars.  We performed
this analysis several times with varying orders for the polynomial
transformations.  Using outlier-resistant statistics, the average
SN~2008ha position from the bootstrapped samples are consistent with
our best estimate.  We use the outlier-resistant scatter of the
resulting positions, 0.44 and 0.50 ACS pixels in the R.A.\ and Dec.\
directions, respectively, which are larger than the nominal
uncertainty from the nominal position, as the uncertainty in the
position of SN~2008ha from the geometric transformation.  To this, we
add in quadrature the statistical uncertainty in the position of
SN~2008ha in the Swope image, which is equivalent to 0.08 ACS pixels,
and the estimated systematic uncertainty in determining the SN
position, which is equivalent to 0.44 ACS pixels.  The total
uncertainty in measuring the position of SN~2008ha in the {\it HST}
images is thus 0.71 and 0.75 ACS pixels (0.037\arcsec\ and
0.039\arcsec) in the R.A.\ and Dec.\ directions, respectively.

Offset by 0.86 ACS pixels, corresponding to 0.043\arcsec, from the SN
position, there is a source clearly detected in the F814W band (${\rm
  S/N} = 6.9$).  This source is marginally detected in the F625W band
(${\rm S/N} = 1.8$), and undetected in the bluer bands.  We label this
source ``S1.''  The uncertainty in the position of this source is
roughly 0.4 ACS pixels.  Adding this in quadrature to the uncertainty
in the position of SN~2008ha, S1 is 1.1~$\sigma$ from the position of
SN~2008ha.  Photometry of S1 yields magnitudes of $<$27.7, $<$27.3,
$28.1 \pm 0.6$, and $26.34 \pm 0.16$~mag in the {\it HST} filters,
from blue to red, respectively, with the limits being 3-$\sigma$
limits.

The ${\rm F625W} - {\rm F814W}$ color of S1 is redder than almost all
nearby stars.  Of the 100 stars closest to the position of SN~2008ha
(including S1), 84 are bluer, and 13 have non-detections in F625W, but
are fainter in F814W such that they could be bluer than S1.  That is,
only 2 of the closest 100 stars are definitely redder than S1.

\begin{figure}
\begin{center}
\epsscale{1.15}
\rotatebox{0}{
\plotone{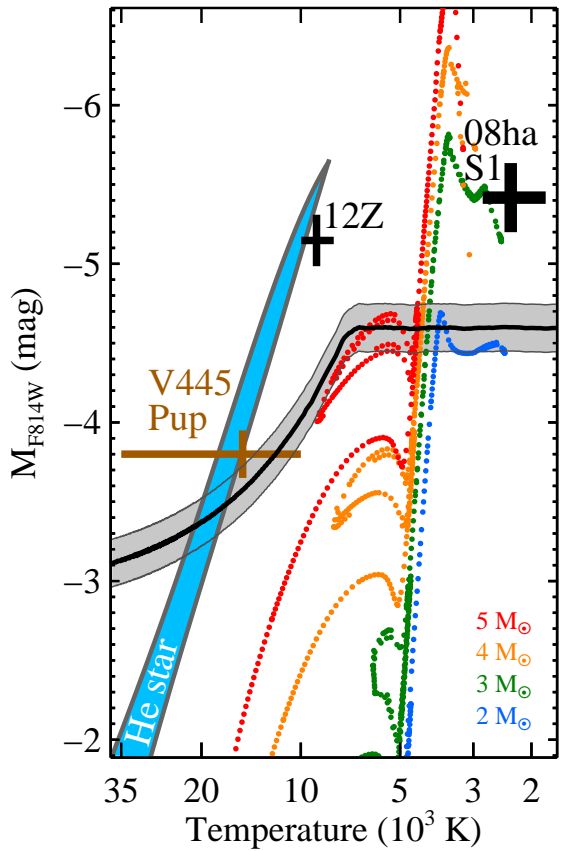}}
\caption{Hertzsprung-Russell diagram for S1 (thick black cross; $2100
  \pm 500$~K), the {\it HST} limits for a blackbody source of a given
  temperature (black line), the SN~2012Z progenitor system (thin black
  cross), and the V445~Pup progenitor system (brown cross; F814W
  magnitude estimated from temperature and $V$-band magnitude).  Also
  plotted are stellar evolution tracks for stars with initial masses
  of 2, 3, 4, and 5~M$_{\sun}$ \citep{Bertelli09} and the region
  predicted for one set of He-star progenitor models from
  \citet{Liu10} (blue region).  The limits shown are the single-band
  3-$\sigma$ limits for black bodies.  The grey band indicates the
  additional 1-$\sigma$ distance modulus uncertainty.  The distance
  uncertainty is included in the uncertainties for S1.}\label{f:hr}
\end{center}
\end{figure}

Assuming our nominal distance modulus and correcting for Milky Way
reddening of $E(B-V) = 0.07$~mag \citep{Schlegel98, Schlafly11}, we
can derive the luminosity of S1.  Fitting a blackbody to the
photometry of S1 (assuming no host-galaxy reddening), we find a
best-fit temperature of $2100 \pm 500$~K and a radius of $1500 \pm
800$~R$_{\sun}$, and a bolometric luminosity of $L_{\rm bol} = 1.5
\times 10^{38}$~erg~s$^{-1}$.  Using this temperature and the F814W
absolute magnitude, we place S1 on a Hertzsprung-Russell (HR) diagram
(Figure~\ref{f:hr}).

Although it is difficult to measure host-galaxy reddening from SN~Iax
colors \citep{Foley13:iax}, SN~2008ha did not appear to have any
host-galaxy reddening \citep{Foley09:08ha}.  The position of the SN in
its host galaxy also makes a substantial amount of dust in the
interstellar medium unlikely.  However, it is possible that there is
newly formed circumstellar dust.  

V445~Pup, a helium nova that ejected \about $10^{-4}$~M$_{\sun}$ of
material formed a significant amount of circumstellar dust 7.5~months
after discovery\citep{Kato08}.  Under the right conditions, SN~2008ha
could have generated a similar amount of dust, which could also
account for the very red colors of S1.  Unfortunately, we cannot
constrain the extinction from newly formed circumstellar dust with our
current observations.  Ignoring the possibility of circumstellar dust,
the colors and luminosity of S1 are consistent with an thermal-pulsing
asymptotic giant branch (TP-AGB) star that had an initial mass of
$>$3~M$_{\sun}$ (see Section~\ref{ss:comp}).

\section{Analysis}\label{s:anal}

Having identified a source that is spatially coincident with
SN~2008ha, we now examine its physical properties in detail.  There
are four scenarios for this source: it is unrelated to SN~2008ha, it
is emission from the SN itself, it is the luminous remnant of the WD
progenitor, or it is the companion star to the WD.

Additionally, we examine constraints on the progenitor system and
remnant if S1 is not associated with SN~2008ha.

\subsection{Chance Coincidence}\label{ss:chance}

There are many bright, resolved stars in our images of UGC~12682.  For
certain positions, there will be a chance superposition with one of
these stars.  As noted above, S1 is 0.86 ACS pixels, equivalently
0.043\arcsec\ and 1.1~$\sigma$, from the position of SN~2008ha.  We
now calculate the probability of such a similar chance alignment.

Our analysis follows that of \citetalias{McCully14:12z}.  Within a $40
\times 320$~pixel (2\arcsec$\times$16\arcsec) box centered on
SN~2008ha, we detect 89 stars with ${\rm S/N} > 3$ in at least one of
our {\it HST} images.  The dimensions of the box were chosen such that
the major axis was roughly perpendicular to the stellar density
gradient, and the density of stars within this box should be roughly
representative of the stellar density at the position of SN~2008ha.

Given our astrometric uncertainty, there is a 0.90\% (7.0\%) chance
that a random position near SN~2008ha would be within 1.1~$\sigma$
(3~$\sigma$) of any detected star (in any of the {\it HST} images),
making a chance alignment somewhat unlikely.  Additionally, there are
only 53 stars within the same area that are brighter than S1; there is
only a 0.54\% (4.2\%) chance of being within 1.1~$\sigma$ (3~$\sigma$)
of a star at least as bright as S1.  Although this latter measurement
is an {\it a posteriori} statistic, it is helpful for determining the
rarity of such stars and what we would have determined for slightly
shallower images.

The likelihood of chance alignment is small, but not sufficiently
small to completely rule out that scenario.  Our above calculation
assumes no spatial clustering of stars, which is incorrect.  However,
it is not possible to precisely determine the necessary correction
with the current data.  Nonetheless, if S1 is not part of the
SN~2008ha progenitor system, but is located nearby because it was born
in the same cluster, this is not a chance coincidence.  In that
scenario, S1 will have a similar age (and composition) as the
SN~2008ha progenitor system.  Future observations could possibly
differentiate these scenarios.

\subsection{Supernova Emission}

\begin{figure}
\begin{center}
\epsscale{1.0}
\rotatebox{90}{
\plotone{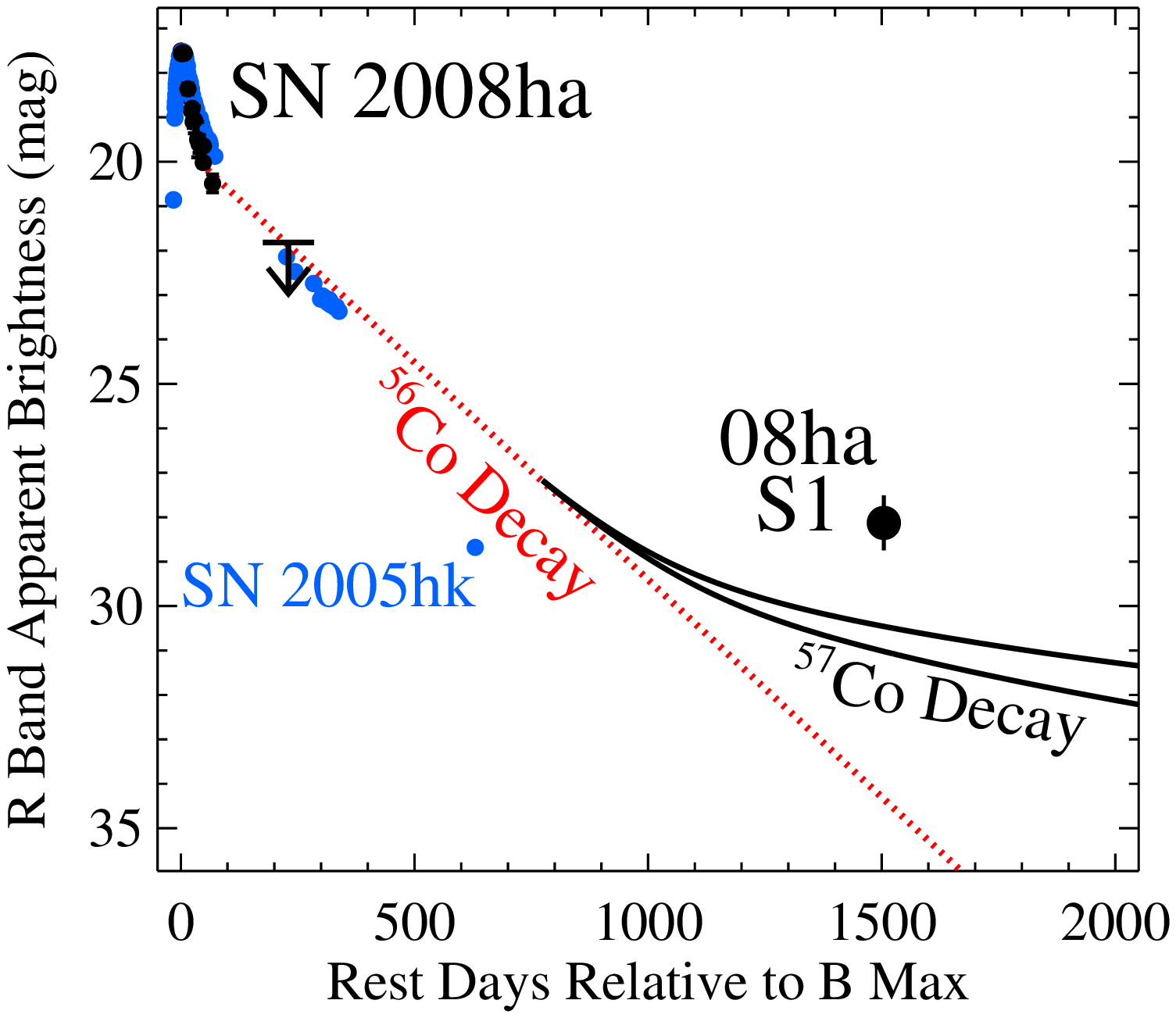}}
\caption{$R$-band light curve of SN~2008ha and photometry of S1
  (converting our blackbody fit to an $R$ band magnitude).  The blue
  points are the $R$-band light curve of SN~2005hk \citep{Phillips07,
    McCully14:iax}, shifted to match the peak brightness of SN~2008ha.
  The red-dotted line is the expected light curve extrapolated from
  the last detection (and is thus an upper limit) if the SN is only
  powered by $^{56}$Co decay.  The black solid lines are the model
  SN~Ia light curves from \citet{Roepke12} that include additional
  radioactive elements.  The two curves represent different explosion
  mechanisms and thus slightly different nucleosynthetic yields.  S1
  is $>$2~mag brighter than the expected emission from SN~2008ha at
  the time of our {\it HST} images.}\label{f:lc}
\end{center}
\end{figure}

SN~2008ha was a very low luminosity event, likely the least luminous
hydrogen-deficient SN yet observed.  It peaked at $V = 17.68$~mag,
corresponding to $M_{V} = -14.17$~mag, and declined quickly with
$\Delta m_{15} (B) = 2.17$~mag \citep{Foley09:08ha}.  SN~2008ha was
last detected at $V = 20.2$~mag 68 days after $B$ maximum
\citep{Valenti09}.  A later non-detection of $R > 22.5$~mag at
230~days after $B$ maximum was consistent with $^{56}$Co decay
extrapolated from the last detection \citep{Foley10:08ha}, but the
true brightness of the SN may have been much fainter.  We reproduce
the SN~2008ha light curve of \citet{Foley10:08ha} in
Figure~\ref{f:lc}, including our observation of S1.  The last
detection and the non-detection are 8.6 and 5.6~mag brighter than S1
(and roughly the same for our limits), respectively.  However, our
observation of S1 was obtained 1505~days after $B$ maximum (in the
rest frame), and extrapolating the $^{56}$Co decay to that time, we
expect the SN to be at least 6.2~mag fainter than S1.

It is expected that SN light curves should fade slower than the
$^{56}$Co decay rate at times $\gtrsim$1000~days after $B$ maximum,
when the $^{57}$Co decay becomes the dominant contribution to the SN
luminosity \citep{Seitenzahl09, Roepke12}.  We place the
\citet{Roepke12} model SN~Ia light curves (scaled in luminosity to
match SN~2008ha) in Figure~\ref{f:lc}.  The models invoke different
explosion mechanisms (either violent merger or delayed detonation),
which result in somewhat different amounts of particular radioactive
nucleosynthetic products and thus slightly different luminosities at
very late times.  The details of the nucleosynthesis may be
particularly important for SN~2008ha, but both model light curves are
much fainter (2.3 and 2.8~mag) than S1 at the time of our observation.
In fact, S1 is brighter than the expected brightness of SN~2008ha at
1000~days, before the other radioactive products are expected to
strongly affect the light curve.  Since the radioactive decay is
exponential, even a doubling of the abundance of these long-lived
radioactive species will not dramatically affect the late-time
luminosity.

Additionally, S1 is more luminous than SN~Iax~2005hk at 630~days after
maximum \citep{McCully14:iax}, after shifting its peak brightness to match
that of SN~2008ha.  Unless the SN faded in an unexpected way (or
rebrightened), S1 is inconsistent with being emission from the SN.
For S1 to be SN emission, it must have faded by an average of
$<$1.4~mag~yr$^{-1}$ from the time of the non-detection.

Any circumstellar dust extinction will only increase the discrepancy
between the expected SN luminosity and that of S1.

A final possibility is that S1 is a light echo of SN~2008ha.  This
possibility is highly unlikely; there is no indication of dust in
front of the SN, S1 is not resolved, and light echoes are usually {\it
  bluer} than the SN.

\subsection{Remnant}

If the explosion fails to unbind the progenitor star, then there will
be a gravitationally bound remnant.  Such an object has never before
been identified, but it is possible that it could be quite luminous.

From early-time spectra, we know that some of the SN~2008ha ejecta is
traveling at speeds much greater than the escape velocity of the WD,
and \citet{Foley10:08ha} estimated that there was 0.3~M$_{\sun}$ of
unbound material.  From mass loss alone, we expect the radius of the
WD to change, but not significantly.  If the progenitor was a
Chandrasekhar-mass WD, it would have lost \about 20\% of its mass in
the explosion and would have expanded to about twice its radius.  At
lower masses, the radius of a WD has a weak dependence on its mass, $R
\propto M^{-1/3}$; a 0.45~M$_{\sun}$ WD losing 0.3~M$_{\sun}$ (67\% of
its mass), would only have its radius increase by 44\%.

Probably more important to the radius and luminosity of any remnant is
the additional energy input directly from the explosion and from
radioactive material left in the star.  If there is a surviving
companion star, then the radius of the remnant is effectively limited
by the Roche radius under normal conditions.  Once the remnant grows
to that radius, mass transfer to the companion will begin.  However,
the conditions 4~years after the explosion may not be close to an
equilibrium; for instance, the remnant may drive a strong opaque wind,
which would increase the effective radius beyond the Roche radius,
creating a common envelope.  This super-Eddington wind could explain
the persistent photosphere with low-velocity P-Cygni profiles seen in
the late-time spectra of SNe~Iax \citep{Jha06:02cx, Foley13:iax,
  McCully14:iax}.

Blackbody fits to the S1 photometry yield a best-fit temperature of
$2100 \pm 500$~K and a radius of $1500 \pm 800$~R$_{\sun}$, both
consistent with a TP-AGB star, although the implied temperature is
significantly lower and the implied radius is significantly large than
most TP-AGB stars and red supergiants (although perhaps consistent
during an extreme mass-loss event).  We caution over-interpretation of
these values; although S1 was strongly detected in one band, it is
only marginally detected in a second band with limits in two other
bands.  Additionally, a blackbody spectrum may not be appropriate for
this source.  Furthermore, circumstellar dust may make S1 appear
redder and fainter than it really is.  However if S1 has minimal
circumstellar reddening and is roughly a blackbody, it must be
somewhat cool and large.

About $2 \times 10^{46}$~erg of energy is needed to expand a
1~M$_{\sun}$ WD to \about 1000~R$_{\sun}$.  There was \about $2 \times
10^{48}$~erg of kinetic energy coupled to the ejecta of SN~2008ha
\citep{Foley09:08ha}.  Therefore, about 1\% of the explosion energy
needs to be injected into the star to expand it to our measured
radius.  Even less energy is needed if only a low-mass envelope
expanded to those radii.  If the star were to do this, the effective
temperature would decrease, and the remnant may appear similar to an
TP-AGB star.

\citet{Jordan12} and \citet{Kromer13} presented single-plume
deflagration explosion models for SNe~Iax.  These models, which do not
burn through the entire WD, left a \about 1~M$_{\sun}$ remnant that
was polluted with nucleosynthetic ash, and particularly some
$^{56}$Ni.  Assuming that there was full $\gamma$-ray trapping in the
remnant, \citet{Kromer13} indicated that the remnant may become more
luminous than the SN at \about 100~days after $B$-band maximum.  The
luminosity of this remnant, however, is tied to the amount of heating
from the explosion and the $^{56}$Ni in the remnant, making the
remnant luminosity only slightly larger than the SN luminosity at late
times.  The remnant luminosity should decrease slower than that of the
SN, but this decay is still larger than observed for S1.  The remnant
was not resolved in their simulation, but \citet{Kromer13} expects the
remnant to expand and be ``puffed up,'' qualitatively consistent with
the large radius inferred for S1.

\subsection{Companion Star}\label{ss:comp}

\citetalias{McCully14:12z} discovered the progenitor system for SN~2012Z,
another SN~Iax, in pre-explosion {\it HST} images.  For that case, it
is impossible for the detected source to be the SN emission or a
remnant.  Although it is possible that this emission came from an
accretion disk around the WD, the best interpretation is that it is
the companion star to the WD that exploded.  For SN~2008ha, with the
deep {\it HST} images obtained after the SN, the interpretation of S1
is not as clear, but it could be a companion star.

Examining Figure~\ref{f:hr}, we see that S1 is near the AGB/red
supergiant portion of the HR diagram, but is redder than model stellar
evolution tracks.  The red color of S1 may be caused by newly formed
circumstellar dust, but we have no constraint on any circumstellar
dust reddening.  Using stellar evolution tracks \citep{Bertelli09}
with a low metallicity ($12 + \log [{\rm O/H}] = 8.16$, as measured
from the nearby \ion{H}{2} region; \citealt{Foley09:08ha}) and
ignoring possible circumstellar reddening, S1 has the temperature and
luminosity of a TP-AGB star.  Because of the scant photometric data
for S1, we use the F814W luminosity to determine the possible initial
mass range.  Since this calculation does not incorporate any color
information, we consider all models consistent to within five times
the photometric uncertainty of the F814W luminosity of S1 to be
consistent.  This limit includes the possibility of circumstellar
reddening of $E(B-V) \le 0.4$~mag ($A_{V} \le 1.2$~mag).  The data are
consistent with a TP-AGB star that is in the midst of dramatic mass
loss.  The luminosity of S1 is consistent with TP-AGB stars with
initial masses of $M > 3$~M$_{\sun}$ (for $Z = 0.004$;
\citealt{Marigo07}).  If there is even more circumstellar reddening,
the temperature could change dramatically, and specific conclusions
about the nature of S1 would need to be adjusted.

This mass range is consistent with having a C/O WD companion, as is
necessary for SN~2008ha \citep{Foley10:08ha}.  Therefore, the TP-AGB
explanation makes S1 a viable candidate for a companion star in the
SN~2008ha progenitor system.

\begin{figure*}
\begin{center}
\epsscale{1.15}
\rotatebox{0}{
\plotone{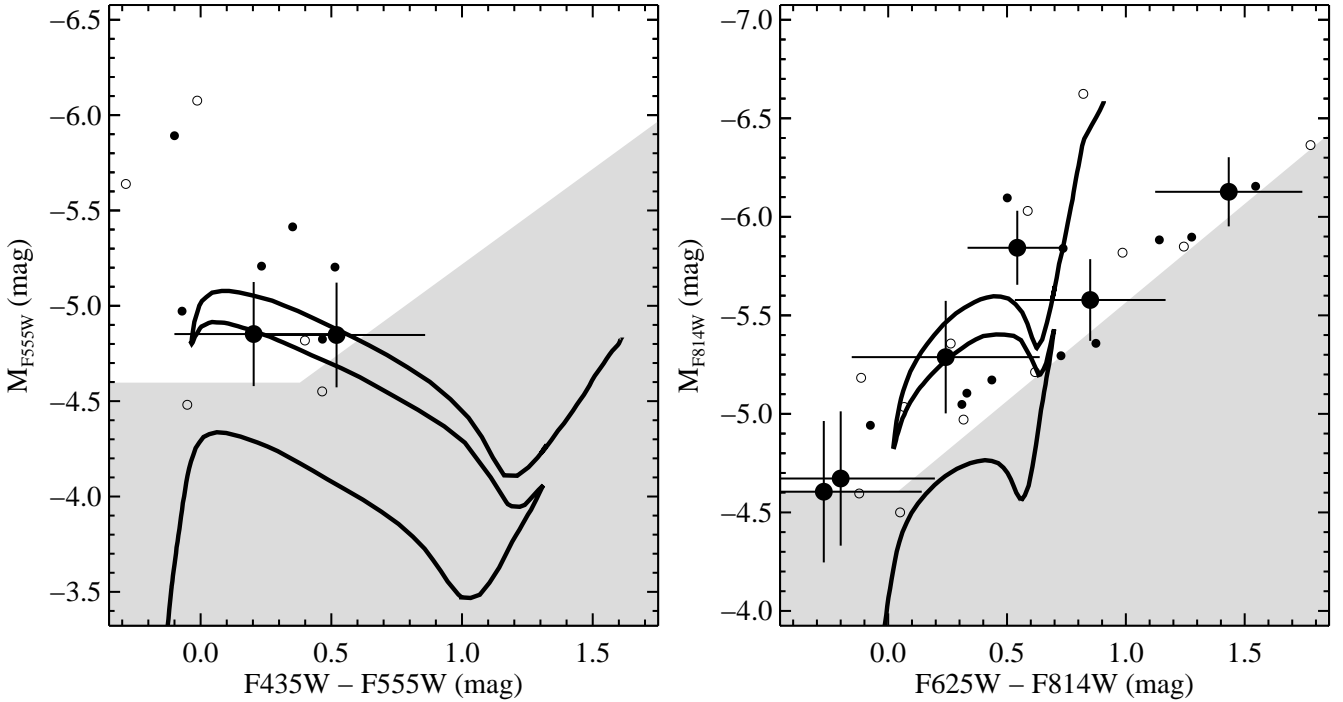}}
\caption{Color-magnitude diagrams for the region near SN~2008ha.  The
  large filled (with error bars), small filled, and open circles
  represent stars within 15, 30, and 45 pixels (0.75, 1.5, and
  2.25\arcsec) of the position of SN~2008ha, respectively.  The grey
  area represents magnitudes below our 3-$\sigma$ detection limit.
  The black curve is a 55~Myr isochrone.}\label{f:iso}
\end{center}
\end{figure*}

SNe~Iax come from relatively young populations \citep{Foley09:08ha,
  Lyman13}, and thus the stars close to the position of SN~2008ha can
constrain the age of the progenitor system.  Fitting isochrones to
nearby stars (\citealt{Bertelli09}; Figure~\ref{f:iso}), we find a
best-fit logarithmic age (in years) of $7.74 \pm 0.09$, corresponding
to $t = 55\err{13}{10}$~Myr.  However, if we exclude one of the four
nearby stars, we determine a 1-$\sigma$ upper limit on the age of
80~Myr.  This age is consistent with our mass range of TP-AGB stars
for S1.

It is possible, although very unlikely, that SN~2008ha had a
massive-star progenitor \citep{Valenti09, Moriya10}.  For this
scenario, the progenitor must have been a very massive star that had
its hydrogen and a significant amount of helium stripped from its
outer layers.  Any such star would have too short a life to allow a 2
-- 6~M$_{\sun}$ companion to evolve to become a TP-AGB star.
Therefore, if S1 is the companion star to the SN~2008ha progenitor, it
is one more reason that its progenitor could not be a massive star.
Furthermore, this scenario requires that the progenitor star, which
would necessarily had a very short lifetime ($<$10~Myr), would be much
younger than nearby stars.

S1 is much redder than the likely companion star for SN~2012Z and the
progenitor of V445~Puppis, a Galactic helium nova \citep{Kato03,
  Woudt09}.  Furthermore, S1 is inconsistent with specific models for
a massive helium-burning star companion \citep{Liu10}, which has been
proposed as the progenitor systems for SNe~Iax \citep{Foley13:iax} and
SN~2012Z in particular \citepalias{McCully14:12z}.  If S1 is intrinsically
as blue as these other stars, but reddened by circumstellar dust, then
S1 must also be much more luminous than these stars.  It is unlikely
that S1 is a reddened, but intrinsically blue, He star.

However, only a small portion of the parameter space for WD/He star
progenitor systems has been probed.  It is possible that S1 is a He
star, but instead of a main-sequence He star, it would be a He red
giant, similar in color and luminosity to R~Cor~Bor.  \citet{Kato08}
examined the stellar evolution of He stars to the red giant branch.
Their approach was similar to that of \citet{Paczynski71} and others.
\citet{Kato08} found that single He stars with $M > 0.8$~M$_{\sun}$
would evolve to become much more luminous and redder, with
luminosities consistent (although typically higher) with S1.  However,
the lowest model temperatures were roughly 4500~K, hotter than S1.
Although this model should be investigated further, especially if it
is found that S1 is reddened, the data are currently inconsistent with
the models.

Although observations of S1 {\it after} the SN are consistent with
being a TP-AGB star, the SN may have affected the observational
properties of any companion star.  For SNe~Ia, it is expected that any
nearby companion will have mass stripped and will be shock heated
\citep[e.g.,][]{Marietta00, Pan12, Shappee13}; as a result, their
luminosity is expected to increase.  \citet{Pan12}, \citet{Pan13}, and
\citet{Liu13} specifically looked at He-star companions, the same
companion star suggested for SN~2012Z \citep{McCully14:12z}.  These
studies found that because of their size, the amount of mass stripping
was much lower for He star companions than main-sequence or TP-AGB
companions, but still suggest that the star will ``puff up'' and
become quite luminous.

None of these studies indicated that the companion star should expand
to \about 1500~R$_{\sun}$ or become as cool as S1.  Moreover, SN~2008ha
had an ejecta mass and kinetic energy one and three orders of
magnitude less than that of a SN~Ia, respectively.  Any effect from
the SN on the companion star should be less than that of a SN~Ia
explosion.  We therefore think that it is unlikely that any companion
star is significantly perturbed, and it is extremely unlikely that S1
is an impacted main-sequence He star, but detailed modeling is
required to see the full effects of this scenario.

As seen in Figure~\ref{f:hr}, S1 has very different physical
parameters from the SN~2012Z companion star.  There are five scenarios
that are consistent with the observations of SNe~2008ha and 2012Z: (1)
S1 is not the companion star of in the SN~2008ha progenitor system and
the blue He-star companion has a luminosity below our current limits,
(2) S1 was a blue He star and the SN significantly altered its
appearance (which we consider unlikely), (3) the blue emission from
the SN~2012Z progenitor system was from the WD accretion disk (a
super-soft source) and the companion star is red and lower luminosity
like S1 (which \citealt{McCully14:12z} consider unlikely), (4) S1 is a He
red giant, simply a slightly more evolved version of the companion
star for SN~2012Z, or (5) SNe~Iax have a variety of progenitor
systems.


\section{Discussion and Conclusions}\label{s:disc}

Using {\it HST} images obtained 4.1 years after the explosion, we
detected a stellar source very close to the position of SN~2008ha,
which we call ``S1.''  We determined that this source is unlikely to
be a chance superposition, but have not ruled out such a coincidence.
We have also shown that the luminosity of S1 is much higher than the
expected luminosity of the SN at the time of the observations, leaving
three possibilities for S1: it is not directly related to SN~2008ha,
it is the remnant of the WD that produced SN~2008ha, or it is the
companion star to the WD that produced SN~2008ha.

Fitting isochrones to stars near the position of SN~2008ha, we have
determined that the stars near SN~2008ha have a likely age of
$<$80~Myr.  This is consistent with the average value determined for
SNe~Iax based on their locations relative to H$\alpha$ emission
\citep[30--50~Myr;][]{Lyman13}.

The photometry of S1 suggests that it is a TP-AGB star with an initial
mass of $>$3~M$_{\sun}$ \citep{Marigo07}.  This mass range is
consistent with being the companion of a C/O WD and inconsistent with
being the companion of a massive star progenitor.

S1 appears to be too cool to be a He red giant.  However, the previous
arguments for a He companion star for SN~Iax progenitor systems, the
properties of the SN~2012Z progenitor system, and the obvious mass
transfer triggering by stellar evolution make this option somewhat
attractive.  Binary evolution models and models of how a SN might
change a He red giant star could improve our understanding of the
plausibility of this option.  Newly formed circumstellar dust may make
such a star appear much redder than its intrinsic color.

Regardless, S1 has observational properties very different from the
progenitor system for SN~2012Z \citepalias{McCully14:12z}.  It is
possible that SNe~2008ha and 2012Z had similar progenitor systems and
S1 has expanded and cooled as the result of the SN impact or that the
emission in the SN~2012Z progenitor system is dominated by emission
from the accretion disk.  However, detailed calculations are required
to test these possibilities.  If S1 is the companion star for the
SN~2008ha progenitor system and its observational properties have not
changed significantly from before the SN explosion, then SNe~Iax may
have come from a variety of progenitor systems.  Perhaps the
significant diversity in the properties of SNe~Iax (SN~2012Z was 40
times as luminous as SN~2008ha at peak), is related to the diversity
of progenitor systems.

It is possible that S1 is an inflated remnant of the WD.  SN~2008ha
ejected only 0.3~M$_{\sun}$ of material \citep{Foley10:08ha} and
therefore likely left a remnant.  Models indicate that this remnant
should be relatively luminous at late times and ``puffed up''
\citep{Jordan12, Kromer13, Fink14}.  S1 has a blackbody radius of
\about 1500~R$_{\sun}$, which is qualitatively consistent with these
models; moreover, the energy required for this expansion is relatively
small compared to the energy of the explosion.

Future observations should be able to further confirm if S1 is the
remnant and may rule it out as a companion star.  The remnant should
vary on years timescales, while it is unlikely that the luminosity of
the companion star will change significantly.  Eventually, we will
build large enough telescopes to obtain a spectrum of S1; if S1 is
indeed the remnant, it will likely have a peculiar spectrum.

Deeper observations at bluer wavelengths will have a better chance of
detecting a helium star (the preferred companion star for SN~2012Z) at
a position more coincident with SN~2008ha than S1.  New {\it HST}
images could also improve our astrometry by both increasing the S/N of
S1 and by measuring additional stars for the astrometric linking.  In
the future, obtaining deep high-resolution images of SNe~Iax will
improve our ability to constrain progenitor systems and remnants
associated with SNe~Iax.

S1 represents potentially the second progenitor system of a
thermonuclear SN detected, after SN~2012Z.  If this is the correct
interpretation, it would complicate the He-star companion scenario of
\citet{Foley13:iax} and require a variety of progenitor systems for
SNe~Iax.  Instead if S1 is the remnant of SN~2008ha, it would be the
first detection of such an object.  Continued monitoring of S1 will
improve our understanding of SN~2008ha and SNe~Iax.

\begin{acknowledgments} 

  {\it Facility:} \facility{Hubble Space Telescope(ACS)}

\bigskip

We thank the anonymous referee for helpful and informed comments.

Based on observations made with the NASA/ESA {\it Hubble Space
  Telescope}, obtained at the Space Telescope Science Institute, which
is operated by the Association of Universities for Research in
Astronomy, Inc., under NASA contract NAS 5--26555.  These observations
are associated with program GO--12999.

This research at Rutgers University was supported through NASA/{\it
  HST} grant GO--12913.01, and National Science Foundation (NSF)
CAREER award AST--0847157 to S.W.J.  M.D.S.\ gratefully acknowledges
generous support provided by the Danish Agency for Science and
Technology and Innovation realized through a Sapere Aude Level 2
grant.  This work was supported by the NSF under grants PHY 11--25915
and AST 11--09174.

\end{acknowledgments}

\bibliographystyle{fapj}
\bibliography{../astro_refs}

\begin{thebibliography}{42}
\expandafter\ifx\csname natexlab\endcsname\relax\def\natexlab#1{#1}\fi

\bibitem[{{Bertelli} {et~al.}(2009){Bertelli}, {Nasi}, {Girardi}, \&
  {Marigo}}]{Bertelli09}
{Bertelli}, G., {Nasi}, E., {Girardi}, L., \& {Marigo}, P. 2009, \aap, 508, 355

\bibitem[{{Chornock} {et~al.}(2006){Chornock}, {Filippenko}, {Branch}, {Foley},
  {Jha}, \& {Li}}]{Chornock06}
{Chornock}, R., {Filippenko}, A.~V., {Branch}, D., {Foley}, R.~J., {Jha}, S.,
  \& {Li}, W. 2006, \pasp, 118, 722

\bibitem[{{Dolphin}(2000)}]{Dolphin00}
{Dolphin}, A.~E. 2000, \pasp, 112, 1383

\bibitem[{{Fink} {et~al.}(2014){Fink}, {Kromer}, {Seitenzahl},
  {Ciaraldi-Schoolmann}, {R{\"o}pke}, {Sim}, {Pakmor}, {Ruiter}, \&
  {Hillebrandt}}]{Fink14}
{Fink}, M., {et~al.} 2014, \mnras, 438, 1762

\bibitem[{{Foley}(2008)}]{Foley08:08ha}
{Foley}, R.~J. 2008, Central Bureau Electronic Telegrams, 1576, 2

\bibitem[{{Foley} {et~al.}(2010{\natexlab{a}}){Foley}, {Brown}, {Rest},
  {Challis}, {Kirshner}, \& {Wood-Vasey}}]{Foley10:08ha}
{Foley}, R.~J., {Brown}, P.~J., {Rest}, A., {Challis}, P.~J., {Kirshner},
  R.~P., \& {Wood-Vasey}, W.~M. 2010{\natexlab{a}}, \apjl, 708, L61

\bibitem[{{Foley} {et~al.}(2009){Foley}, {Chornock}, {Filippenko},
  {Ganeshalingam}, {Kirshner}, {Li}, {Cenko}, {Challis}, {Friedman}, {Modjaz},
  {Silverman}, \& {Wood-Vasey}}]{Foley09:08ha}
{Foley}, R.~J., {et~al.} 2009, \aj, 138, 376

\bibitem[{{Foley} {et~al.}(2010{\natexlab{b}}){Foley}, {Rest}, {Stritzinger},
  {Pignata}, {Anderson}, {Hamuy}, {Morrell}, {Phillips}, \&
  {Salgado}}]{Foley10:08ge}
------. 2010{\natexlab{b}}, \aj, 140, 1321

\bibitem[{{Foley} {et~al.}(2013){Foley}, {Challis}, {Chornock},
  {Ganeshalingam}, {Li}, {Marion}, {Morrell}, {Pignata}, {Stritzinger},
  {Silverman}, {Wang}, {Anderson}, {Filippenko}, {Freedman}, {Hamuy}, {Jha},
  {Kirshner}, {McCully}, {Persson}, {Phillips}, {Reichart}, \&
  {Soderberg}}]{Foley13:iax}
------. 2013, \apj, 767, 57

\bibitem[{{Iben} \& {Tutukov}(1991)}]{Iben91}
{Iben}, Jr., I., \& {Tutukov}, A.~V. 1991, \apj, 370, 615

\bibitem[{{Jha} {et~al.}(2006){Jha}, {Branch}, {Chornock}, {Foley}, {Li},
  {Swift}, {Casebeer}, \& {Filippenko}}]{Jha06:02cx}
{Jha}, S., {Branch}, D., {Chornock}, R., {Foley}, R.~J., {Li}, W., {Swift},
  B.~J., {Casebeer}, D., \& {Filippenko}, A.~V. 2006, \aj, 132, 189

\bibitem[{{Jordan} {et~al.}(2012){Jordan}, {Perets}, {Fisher}, \& {van
  Rossum}}]{Jordan12}
{Jordan}, IV, G.~C., {Perets}, H.~B., {Fisher}, R.~T., \& {van Rossum}, D.~R.
  2012, \apjl, 761, L23

\bibitem[{{Kato} \& {Hachisu}(2003)}]{Kato03}
{Kato}, M., \& {Hachisu}, I. 2003, \apjl, 598, L107

\bibitem[{{Kato} {et~al.}(2008){Kato}, {Hachisu}, {Kiyota}, \& {Saio}}]{Kato08}
{Kato}, M., {Hachisu}, I., {Kiyota}, S., \& {Saio}, H. 2008, \apj, 684, 1366

\bibitem[{{Kromer} {et~al.}(2013){Kromer}, {Fink}, {Stanishev}, {Taubenberger},
  {Ciaraldi-Schoolman}, {Pakmor}, {R{\"o}pke}, {Ruiter}, {Seitenzahl}, {Sim},
  {Blanc}, {Elias-Rosa}, \& {Hillebrandt}}]{Kromer13}
{Kromer}, M., {et~al.} 2013, \mnras, 429, 2287

\bibitem[{{Li} {et~al.}(2003){Li}, {Filippenko}, {Chornock}, {Berger},
  {Berlind}, {Calkins}, {Challis}, {Fassnacht}, {Jha}, {Kirshner}, {Matheson},
  {Sargent}, {Simcoe}, {Smith}, \& {Squires}}]{Li03:02cx}
{Li}, W., {et~al.} 2003, \pasp, 115, 453

\bibitem[{{Liu} {et~al.}(2010){Liu}, {Chen}, {Wang}, \& {Han}}]{Liu10}
{Liu}, W.-M., {Chen}, W.-C., {Wang}, B., \& {Han}, Z.~W. 2010, \aap, 523, A3

\bibitem[{{Liu} {et~al.}(2013){Liu}, {Pakmor}, {Seitenzahl}, {Hillebrandt},
  {Kromer}, {R{\"o}pke}, {Edelmann}, {Taubenberger}, {Maeda}, {Wang}, \&
  {Han}}]{Liu13}
{Liu}, Z.-W., {et~al.} 2013, \apj, 774, 37

\bibitem[{{Lyman} {et~al.}(2013){Lyman}, {James}, {Perets}, {Anderson},
  {Gal-Yam}, {Mazzali}, \& {Percival}}]{Lyman13}
{Lyman}, J.~D., {James}, P.~A., {Perets}, H.~B., {Anderson}, J.~P., {Gal-Yam},
  A., {Mazzali}, P., \& {Percival}, S.~M. 2013, \mnras, 434, 527

\bibitem[{{Maoz} {et~al.}(2013){Maoz}, {Mannucci}, \& {Nelemans}}]{Maoz13}
{Maoz}, D., {Mannucci}, F., \& {Nelemans}, G. 2013, ArXiv e-prints, 1312.0628

\bibitem[{{Marietta} {et~al.}(2000){Marietta}, {Burrows}, \&
  {Fryxell}}]{Marietta00}
{Marietta}, E., {Burrows}, A., \& {Fryxell}, B. 2000, \apjs, 128, 615

\bibitem[{{Marigo} \& {Girardi}(2007)}]{Marigo07}
{Marigo}, P., \& {Girardi}, L. 2007, \aap, 469, 239

\bibitem[{{Maund} {et~al.}(2010){Maund}, {Wheeler}, {Wang}, {Baade},
  {Clocchiatti}, {Patat}, {H{\"o}flich}, {Quinn}, \& {Zelaya}}]{Maund10:05hk}
{Maund}, J.~R., {et~al.} 2010, \apj, 722, 1162

\bibitem[{{McClelland} {et~al.}(2010){McClelland}, {Garnavich}, {Galbany},
  {Miquel}, {Foley}, {Filippenko}, {Bassett}, {Wheeler}, {Goobar}, {Jha},
  {Sako}, {Frieman}, {Sollerman}, {Vinko}, \& {Schneider}}]{McClelland10}
{McClelland}, C.~M., {et~al.} 2010, \apj, 720, 704

\bibitem[{{McCully} {et~al.}(2014{\natexlab{a}}){McCully}, {Jha}, {Foley},
  {Chornock}, {Holtzman}, {Balam}, {Branch}, {Filippenko}, {Frieman}, {Fynbo},
  {Galbany}, {Ganeshalingam}, {Garnavich}, {Graham}, {Hsiao}, {Leloudas},
  {Leonard}, {Li}, {Riess}, {Sako}, {Schneider}, {Silverman}, {Sollerman},
  {Steele}, {Thomas}, {Wheeler}, \& {Zheng}}]{McCully14:iax}
{McCully}, C., {et~al.} 2014{\natexlab{a}}, \apj, 786, 134

\bibitem[{{McCully} {et~al.}(2014{\natexlab{b}}){McCully}, {Jha}, {Foley}, 
  {et~al.}}]{McCully14:12z}
{McCully}, C., {Jha}, S.~W., {Foley}, R.~J., \& {et~al.} 2014{\natexlab{b}},
  \nat, in press.

\bibitem[{{Milne} {et~al.}(2010){Milne}, {Brown}, {Roming}, {Holland},
  {Immler}, {Filippenko}, {Ganeshalingam}, {Li}, {Stritzinger}, {Phillips},
  {Hicken}, {Kirshner}, {Challis}, {Mazzali}, {Schmidt}, {Bufano}, {Gehrels},
  \& {Vanden Berk}}]{Milne10}
{Milne}, P.~A., {et~al.} 2010, \apj, 721, 1627

\bibitem[{{Moriya} {et~al.}(2010){Moriya}, {Tominaga}, {Tanaka}, {Nomoto},
  {Sauer}, {Mazzali}, {Maeda}, \& {Suzuki}}]{Moriya10}
{Moriya}, T., {Tominaga}, N., {Tanaka}, M., {Nomoto}, K., {Sauer}, D.~N.,
  {Mazzali}, P.~A., {Maeda}, K., \& {Suzuki}, T. 2010, \apj, 719, 1445

\bibitem[{{Narayan} {et~al.}(2011){Narayan}, {Foley}, {Berger}, {Botticella},
  {Chornock}, {Huber}, {Rest}, {Scolnic}, {Smartt}, {Valenti}, {Soderberg},
  {Burgett}, {Chambers}, {Flewelling}, {Gates}, {Grav}, {Kaiser}, {Kirshner},
  {Magnier}, {Morgan}, {Price}, {Riess}, {Stubbs}, {Sweeney}, {Tonry},
  {Wainscoat}, {Waters}, \& {Wood-Vasey}}]{Narayan11}
{Narayan}, G., {et~al.} 2011, \apjl, 731, L11+

\bibitem[{{Paczy{\'n}ski}(1971)}]{Paczynski71}
{Paczy{\'n}ski}, B. 1971, \actaa, 21, 1

\bibitem[{{Pan} {et~al.}(2012){Pan}, {Ricker}, \& {Taam}}]{Pan12}
{Pan}, K.-C., {Ricker}, P.~M., \& {Taam}, R.~E. 2012, \apj, 750, 151

\bibitem[{{Pan} {et~al.}(2013){Pan}, {Ricker}, \& {Taam}}]{Pan13}
------. 2013, \apj, 773, 49

\bibitem[{{Phillips} {et~al.}(2007){Phillips}, {Li}, {Frieman}, {Blinnikov},
  {DePoy}, {Prieto}, {Milne}, {Contreras}, {Folatelli}, {Morrell}, {Hamuy},
  {Suntzeff}, {Roth}, {Gonz{\'a}lez}, {Krzeminski}, {Filippenko}, {Freedman},
  {Chornock}, {Jha}, {Madore}, {Persson}, {Burns}, {Wyatt}, {Murphy}, {Foley},
  {Ganeshalingam}, {Serduke}, {Krisciunas}, {Bassett}, {Becker}, {Dilday},
  {Eastman}, {Garnavich}, {Holtzman}, {Kessler}, {Lampeitl}, {Marriner},
  {Frank}, {Marshall}, {Miknaitis}, {Sako}, {Schneider}, {van der Heyden}, \&
  {Yasuda}}]{Phillips07}
{Phillips}, M.~M., {et~al.} 2007, \pasp, 119, 360

\bibitem[{{Puckett} {et~al.}(2008){Puckett}, {Moore}, {Newton}, \&
  {Orff}}]{Puckett08}
{Puckett}, T., {Moore}, C., {Newton}, J., \& {Orff}, T. 2008, Central Bureau
  Electronic Telegrams, 1567, 1

\bibitem[{{R{\"o}pke} {et~al.}(2012){R{\"o}pke}, {Kromer}, {Seitenzahl},
  {Pakmor}, {Sim}, {Taubenberger}, {Ciaraldi-Schoolmann}, {Hillebrandt},
  {Aldering}, {Antilogus}, {Baltay}, {Benitez-Herrera}, {Bongard}, {Buton},
  {Canto}, {Cellier-Holzem}, {Childress}, {Chotard}, {Copin}, {Fakhouri},
  {Fink}, {Fouchez}, {Gangler}, {Guy}, {Hachinger}, {Hsiao}, {Chen},
  {Kerschhaggl}, {Kowalski}, {Nugent}, {Paech}, {Pain}, {Pecontal}, {Pereira},
  {Perlmutter}, {Rabinowitz}, {Rigault}, {Runge}, {Saunders}, {Smadja},
  {Suzuki}, {Tao}, {Thomas}, {Tilquin}, \& {Wu}}]{Roepke12}
{R{\"o}pke}, F.~K., {et~al.} 2012, \apjl, 750, L19

\bibitem[{{Schlafly} \& {Finkbeiner}(2011)}]{Schlafly11}
{Schlafly}, E.~F., \& {Finkbeiner}, D.~P. 2011, \apj, 737, 103

\bibitem[{{Schlegel} {et~al.}(1998){Schlegel}, {Finkbeiner}, \&
  {Davis}}]{Schlegel98}
{Schlegel}, D.~J., {Finkbeiner}, D.~P., \& {Davis}, M. 1998, \apj, 500, 525

\bibitem[{{Seitenzahl} {et~al.}(2009){Seitenzahl}, {Taubenberger}, \&
  {Sim}}]{Seitenzahl09}
{Seitenzahl}, I.~R., {Taubenberger}, S., \& {Sim}, S.~A. 2009, \mnras, 400, 531

\bibitem[{{Shappee} {et~al.}(2013){Shappee}, {Stanek}, {Pogge}, \&
  {Garnavich}}]{Shappee13}
{Shappee}, B.~J., {Stanek}, K.~Z., {Pogge}, R.~W., \& {Garnavich}, P.~M. 2013,
  \apjl, 762, L5

\bibitem[{{Stritzinger} {et~al.}(2014){Stritzinger}, {Hsiao}, {Valenti},
  {Taddia}, {Rivera-Thorsen}, {Leloudas}, {Maeda}, {Pastorello}, {Phillips},
  {Pignata}, {Baron}, {Burns}, {Contreras}, {Folatelli}, {Hamuy},
  {H{\"o}flich}, {Morrell}, {Prieto}, {Benetti}, {Campillay}, {Haislip},
  {LaClutze}, {Moore}, \& {Reichart}}]{Stritzinger14}
{Stritzinger}, M.~D., {et~al.} 2014, \aap, 561, A146

\bibitem[{{Valenti} {et~al.}(2009){Valenti}, {Pastorello}, {Cappellaro},
  {Benetti}, {Mazzali}, {Manteca}, {Taubenberger}, {Elias-Rosa}, {Ferrando},
  {Harutyunyan}, {Hentunen}, {Nissinen}, {Pian}, {Turatto}, {Zampieri}, \&
  {Smartt}}]{Valenti09}
{Valenti}, S., {et~al.} 2009, \nat, 459, 674

\bibitem[{{Woudt} {et~al.}(2009){Woudt}, {Steeghs}, {Karovska}, {Warner},
  {Groot}, {Nelemans}, {Roelofs}, {Marsh}, {Nagayama}, {Smits}, \&
  {O'Brien}}]{Woudt09}
{Woudt}, P.~A., {et~al.} 2009, \apj, 706, 738

\end{thebibliography}


\end{document}